# Volumetric modulated arc therapy or step-shoot IMRT? A 4D dosimetry study of motion effect in lung SBRT using a dynamic virtual patient model

Tianjun Ma[1], Bingqi Guo[1], Salim Balik[1], Peng Qi[1], Anthony Magnelli[1], Gregory M M. Videtic[1], Kevin L Stephans[1], Tingliang Zhuang[1,2,*]

[1]Department of Radiation Oncology, Taussig Cancer Center, The Cleveland Clinic Foundation, 10201 Carnegie Ave, Cleveland, OH

[2]Department of Radiation Oncology, University of Texas Southwestern Medical Center, 5323 Harry Hines Blvd., Dallas, TX

**\*CORRESPONDENCE:**

Tingliang Zhuang, PhD

Department of Radiation Oncology, University of Texas Southwestern Medical Center, 5323 Harry Hines Blvd., Dallas, TX 75390, USA

# Abstract

**Purpose:** To investigate the impact of delivery techniques and planning parameters on interplay effect in lung SBRT.

**Methods:** A dynamic virtual patient model containing normal structures and a tumor with adjustable sizes, locations, and 3D breathing motion was utilized. SBRT plans were developed using both step-and-shoot IMRT and VMAT with different planning parameters (energy, isocenter location, PTV margin, and PTV dose heterogeneity). 4D doses were calculated by simulating synchronized delivery of SBRT to the virtual patient model with random initial positions of tumor motion. The expected dose (average) and the standard deviation of the 4D doses were obtained. The relative difference between the expected GTV minimal/mean ($GTV_{Min}$/$GTV_{Mean}$) dose and the planned $ITV_{Min}$/$ITV_{Mean}$ dose (denoted by %E/P), and between the $GTV_{Min}$ and the prescription dose ($D_{Rx}$) were computed.

**Results:** The %E/P for $GTV_{Mean}$ was significantly lower for IMRT than VMAT (0.5%±7.7% v.s. 3.5%±5.0%, p=0.04). The expected $GTV_{Min}$ was lower than $D_{Rx}$ in 9.4% of all IMRT plans versus 3.1% in VMAT. The worst-case scenario, 4D $GTV_{Min}$ was 14.1% lower than the $ITV_{Min}$.
Choices of PTV margin or dose heterogeneity to be achieved in PTV can result in significant difference (p<0.05) in motion interplay depending on delivery techniques.

**Conclusion:** Motion interplay may cause the expected $GTV_{Min}$ to be less than the planned ITV minimal dose and $D_{Rx}$ for both IMRT and VMAT plans. The differences between the expected GTV dose and the ITV dose depended on the delivery technique and planning parameters. Overall, VMAT is less prone to motion interplay than IMRT.

# Introduction

Stereotactic body radiation therapy (SBRT) has become a standard of care for treating early-stage non-small cell lung cancer (NSCLC) considered surgically inoperable [1–7]. Both step-and-shoot intensity-modulated radiotherapy (IMRT) and volumetric modulated arc therapy (VMAT) have been used in lung SBRT to improve conformity to targets and normal tissue sparing [8,9].

Respiratory breathing motion presents challenges for radiation treatment of lung tumors [10–14]. A concern in using highly modulated radiotherapy such as IMRT or VMAT for thoracic targets has been the potential under- or over-dosing caused by the interplay between multi-leaf collimators (MLCs) motion and the tumor motion during the treatment. Yu et al. calculated the interplay effect for the sliding-window IMRT and found a dose variation greater than 50% of the desired beam intensity [15].

The motion interplay effect for lung IMRT with conventional dose fractionation has been extensively studied [16–22]. A point dose variation for a single field can be quite large (>30%) [16,17], but is reduced to 2-3% after averaging over a course of 30-fraction treatment [16]. In the era of SBRT, Ong et al. conducted film measurements on 20 VMAT patient plans using phantom and found that even for excessive MLC modulations and 25 mm tumor motion amplitude, the measured dose agreed with the planned dose using gamma analysis [23]. Limited impact of motion interplay was also observed by Rao et al. in a simulation study of 10 patients, in which less than 1% of prescription dose was found for the target in both IMRT and VMAT techniques [24]. Similar studies also demonstrated negligible motion interplay effect for SBRT [25–29]. However, Zhao *et al.* demonstrated the interplay effect may not be negligible in a gated SBRT [30]. Other studies found the number of breaths per treatment [31] and ITV to PTV margin with VMAT [32] may have an impact on the dose interplay effect.

That said, measurements performed in a phantom does not represent the complexity of human anatomy. Simulations performed on clinical patient plans may be limited by tumor motion patterns. A potential solution to these limitations was to model the patient anatomy and various tumor motion patterns using a dynamic virtual patient model. Additionally, the two forms of intensity-modulated radiation therapy: static gantry angle step-and-shoot IMRT and VMAT represented different patterns of MLC leaf trajectory versus time (piece-wise constant in IMRT vs. continuous in VMAT) which could result in differences in motion interplay effect. Furthermore, different setup margins for PTV and different beam energies can be used in treatment planning, which may also lead to distinct motion interplay effect. There is a lack of a systematic study regarding the impact of delivery techniques, planning parameters and plan modulations for representative tumor motion patterns on motion interplay in a patient-like geometry. Such a study may reveal the sensitivity of a plan to motion interplay for different combinations of delivery technique and planning parameters, such as beam energy, PTV margin, dose heterogeneity in PTV[33]. With this in mind, we investigated the motion interplay effect in SBRT of lung tumors using a dynamic virtual patient model built upon 4DCT images of a lung tumor patient [34]. The model simulates lung patients with different tumor sizes, tumor locations, and tumor motion patterns. The simulated patients are planned using both IMRT and VMAT techniques with different combinations of planning parameters such as beam energy, isocenter location, PTV margin, and PTV dose heterogeneity. A 4D dosimetry technique based on synchronization of the dynamic virtual patient model with the delivery beam was developed to calculate the motion interplay effect. Our goal is to recommend treatment techniques and/or planning parameters that may result in SBRT lung plans that are less sensitive to motion interplay.

# Material and methods
## 1. Virtual Patient Model

A dynamic virtual patient model previously developed based on 4DCT of a lung patient was adapted for this study[34]. The model includes 4D non-uniform rational B-spline (NURBS) models of the tumor and normal structures in the thorax and abdomen region, which was built upon the 4DCT and contours of a lung patient. It is capable of simulating various tumor sizes, densities, locations, and regular or irregular breathing motion patterns. Both the tumor and normal structures were deformed along with the breathing motion.

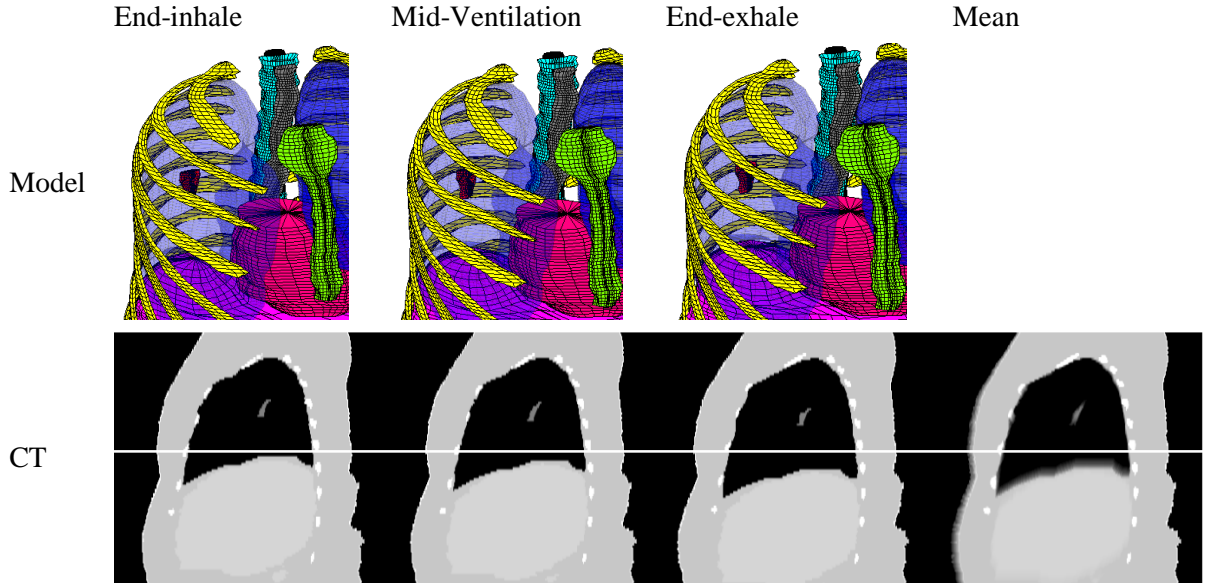

**Figure** 1. Demonstration of the virtual patient model (first row) and sagittal CT slices (second row) at end-inhale, mid-ventilation and end-exhale, and AIP CT for a 2 cm tumor located in the center of the right lung, moving with motion amplitude of 1cm.

In this study, periodic 3D breathing motion with a period of 4s was simulated by repeating one cycle of a patient breathing curve. The amplitudes of the tumor anterior-posterior and left-right motion were fixed at 0.5 cm and 0.2 cm respectively. The amplitude of the tumor superior-inferior motion was scalable ranging from 1 cm to 3 cm. To simulate tumors with different sizes, the tumor contour (GTV) from a lung patient was linearly scaled in all three dimensions from 0.5 cm to 2 cm. In this paper, the tumor size was specified by the maximum dimension of the free-breathing GTV. The virtual patient model can be converted to a reference CT (defined as the 3D CT at end-inhale in this study) and a time-series of individual CT sets (*iCTs*) with a defined time resolution (0.4s in this study). At the same time, the deformation maps from each *iCT* to reference CT were obtained. In this paper, the set of *iCTs* was generated at *n* equally spaced positions along the breathing curve. The number *n* was chosen to be 10 as a balance between accuracy and computational time. The end-inhale CT was selected as the reference CT. The tumor and normal structures in each static *iCT* are assigned with fixed Hounsfield unit (HU) values (same HU value for the entire structure) listed in **Table 1**. The average intensity projection (AIP) CT ($CT_{AIP}$) generated from *iCTs* was used for treatment planning. **Figure** 1 demonstrates the model and corresponding *iCT* at end-inhale, mid-ventilation and end-exhale for a 5mm tumor with 1cm breathing motion amplitude. The $CT_{AIP}$ of all the *iCTs* is also shown in **Figure** 1.

**Table 1**. Assigned HU value to convert virtual patient model to *iCTs*

| Structure | HU value |
|---|---|
| Tumor | -250 |
| Lung | -750 |
| Spinal cord | 13 |
| Rib | 700 |
| Heart | 56 |
| Sternum | 385 |
| Vertebra | 514 |
| Esophagus | 41 |
| Body/Muscle | 0 |
| Skin | 60 |

**1.A. Lung Tumor Modes: varying tumor sizes, locations and motion amplitudes**

Four tumor models, depicted in **Figure** 2, with various GTV sizes, locations, and motion amplitudes were simulated in this study: (1) island lesion, 5 mm GTV size and 3 cm tumor motion; (2) island lesion, 5 mm GTV size and 1 cm tumor motion; (3) island lesion, 2 cm GTV size and 1 cm tumor motion, and (4) central lesion, 2 cm GTV size and 1 cm tumor motion.

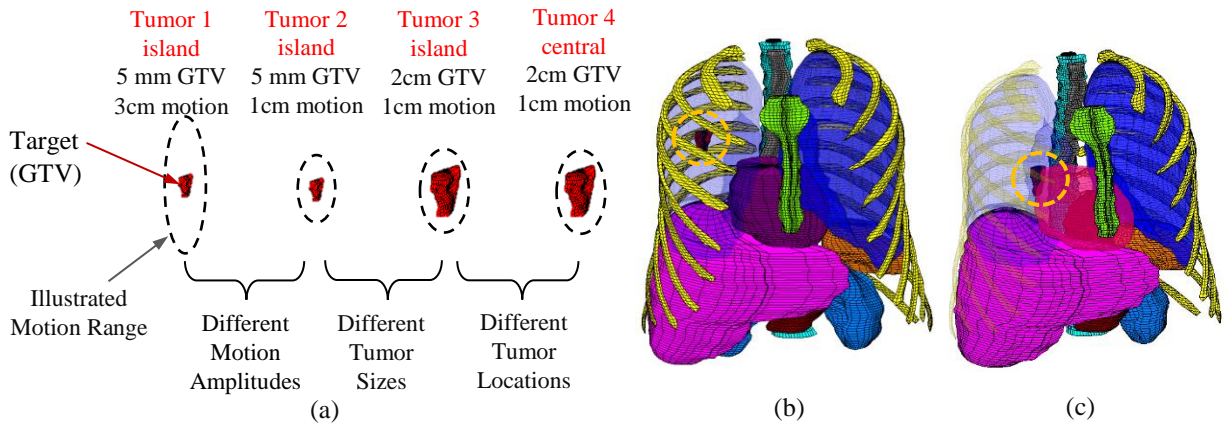

**Figure** 2. Lung tumor model demonstration (a) diagram of GTV with different motion amplitudes, sizes, and locations, (b) island- and (c) central-tumor in the virtual patient model

**2. Treatment Planning**

For each tumor model, 32 plans were designed on the $CT_{AIP}$ using different combinations of delivery techniques (step-and-shoot IMRT and VMAT) and planning parameters (beam energy / isocenter location / PTV margin / PTV dose heterogeneity as shown in ). A total number of 128 plans (64 IMRT and 64 VMAT) were made in this study. The treatment planning system is Pinnacle$^3$ version 9.10 (Philips Radiation Oncology System, Fitchburg, WI). All IMRT plans have seven step-and-shoot beams with 20º gantry angle spacing on the ipsilateral (right) side of the body. The total number of segments for each IMRT plan was chosen to be 20. VMAT plans consist of two 200º partial arcs starting from the posterior direction with collimator angles at 20° and 340°. The planning goals were following the dose constraints in RTOG 0813 but with a prescription of 40 Gy in five fractions. Note, this is a non-clinical schedule proposed in the RTOG 0813 [35]. All the IMRT and VMAT plans were further divided into eight subgroups. Within each subgroup, all the plans share one common planning parameter as listed in **Table 2**. Each subgroup was labeled by its common planning parameter. For example, the plans in the 5-mm IMRT subgroup consist of plans with 5 mm ITV to PTV margin but various beam energies, isocenter locations and PTV dose distributions.

**Table 2**. Planning parameters investigated in this study for each tumor model

|  | Planning parameters [§] | |
|---|---|---|
| Beam Energy | 6 MV FFF | 10 MV FFF |
| isocenter location | COM of PTV (C-ISO) | Off-COM of PTV [†] (O-ISO) |
| ITV to PTV Margin | 5 mm | 3 mm |
| PTV dose distribution | PTV max dose < 110% (homogeneous PTV dose) | 110%<PTV max dose < 120% (heterogeneous PTV dose) |

[§] $2^4$=16 combinations of planning parameters for IMRT and VMAT planning for each tumor model
[†] The Off-center of mass (COM) point is 5 mm away in three orthogonal directions from the PTV COM

### 3. 4D Dose Calculation

To calculate 4D doses for each IMRT/VMAT plan for a given tumor model and motion pattern, the delivery of the IMRT/VMAT plan was emulated using an in-house MATLAB program, where the entire plan was parsed into SubPlans to each *iCT*. SubDoses from the SubPlans were calculated using Pinnacle treatment planning system. 4D Dose was obtained in MATLAB as summation of each SubDose deformed to $CT_{Ref}$. The steps are presented as the pseudo-code in Algorithm 1, and explained below:

```
Input: tumor model parameters and motion pattern
1 begin
2    CT_Ref, 10 iCTs and corresponding deformation maps to CT_Ref, and CT_AIP ← generated per
         given tumor model and motion pattern (in MATLAB)
3    Plan_ini ← generate initial treatment plan on the CT_AIP (in Pinnacle)
4    SubPlans ← generated based on the Plan_ini, motion pattern and initial positions of tumor
         motion (in MATLAB)
5    for i ← 1 to 10 ∈ initial positions of tumor motion do
6        for j ← 1 to 10 ∈ iCTs do
7            SubDose(i,j) ← calculate dose of each SubPlan(i,j) on iCT(j) (in Pinnacle)
8            SubDose_ref(i,j) ← deform the SubDose(i,j) back onto CT_Ref (in MATLAB)
9        end
10       4D dose(i) ← sum of SubDose_ref(i,j) (in MATLAB)
11   end
12 end
Result: 4D doses for different initial positions of tumor motion
```

Algorithm 1. Pseudo-code workflow for 4D dose calculation

To generate *SubPlans* per *iCT* according to the motion pattern and initial position of tumor motion, the IMRT or VMAT plan was divided into *N* MU segments (*N* = 400 in this study) with equal MUs, as shown in **Figure** 3(a). The program simulates the delivery process to determine the time-stamp for each MU segment that contains information of corresponding gantry angle, jaws and MLC leave positions. Depending on the initial position of tumor motion, all the MU segments belonging to the same *iCT* were resembled as a *SubPlan* (Line 4 in Algorithm 1). In this paper, 10 *iCT* were generated for each tumor model, and 10 *SubPlan* generated from each initial IMRT/VMAT plan for a given initial position of tumor motion were calculated on the corresponding *iCTs*. To calculate the dose for each *SubPlan* on the corresponding *iCT*, all the MU segments were treated as static beams. The *SubDoses* were then deformed back to the $CT_{Ref}$ based on the deformation map generated from the model (Line 8 in Algorithm 1) [34]. Finally, the 4D dose was obtained as the summation of 10 deformed $SubDoses_{Ref}$ on $CT_{Ref}$ (Line 10 in Algorithm 1) as demonstrated in **Figure** 3(b). The steps from line 5 to line 11 in Algorithm 1 were repeated to acquire all the *4D doses* for different initial positions of tumor motion.

## 4. Data Analysis

To quantify the dosimetric effect of motion interplay, we first obtained the mean or minimal (Min) GTV dose from each of the ten 4D doses, noted as $D_i$. The expected GTV mean or minimal dose was calculated as the average $\bar{D} = \sum_{i=1}^{i=10} D_i/10$ and the standard deviation of 4D GTV mean or minimal dose was computed and normalized to the prescription dose.

The expected GTV mean and minimal dose were compared to the corresponding mean and minimal dose of ITV from the initial plan on the $CT_{AIP}$. The percentage difference of expected GTV dose over planned ITV dose (%E/P) was computed for all the plans as follows:

$\%E/P = (\bar{D}_{GTV} - D_{ITV})/D_{ITV} \times 100\%$ ,

where $D_{ITV}$ represented either planned ITV mean or minimal dose, and $\bar{D}_{GTV}$ is the corresponding expected GTV mean or minimal dose. A negative value of %E/P indicates the expected GTV dose after considering the motion interplay effect would be less than the planned ITV dose.

To evaluate which delivery technique or planning parameter may be selected to minimize the impact of motion interplay effect, the following comparisons were conducted: (1) the entire IMRT group to the entire VMAT group; (2) for each subgroup, compare IMRT plans to its corresponding VMAT plans (e.g. 5 mm PTV margin IMRT plans to 5 mm PTV margin VMAT plans); and (3) for each delivery technique (IMRT or VMAT), comparing two plan subgroups with the common planning parameter in the same category.(e.g. 5 mm PTV margin IMRT plans to 3mm PTV margin IMRT plans). The non-parametric, Wilcoxon signed-rank test, was utilized to test the difference of dosimetric impact of motion interplay between different planning groups/subgroups. The differences were considered statistically significant if *p* values were equal to or less than 0.05. All the comparisons were based on the dose for a single fraction.

In addition, the number of plans in which the expected GTV minimal dose satisfied the following two criteria was recorded:

(a) $\%E/P \geq 0$, or

(b) $\%E/P > -3\%$ and expected GTV minimal dose > prescription dose.

In criterion (a), the expected GTV minimal dose was greater than the planned ITV minimal dose and was defined as "no underdose". In criterion (b), the expected GTV minimal dose can be slightly less than the planned ITV minimal dose but no less than the prescription dose and was defined as "borderline no underdose" in this paper.

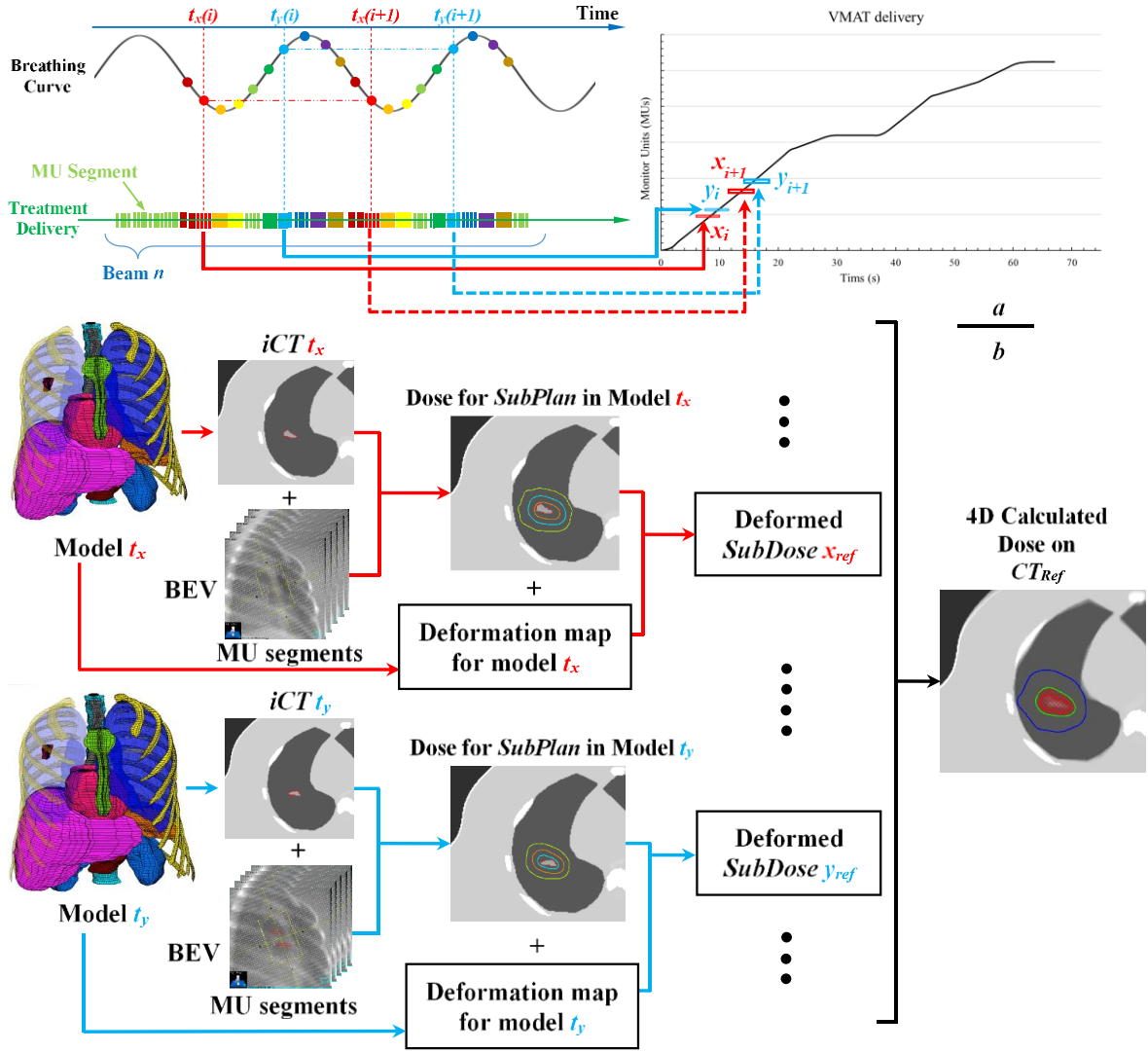

**Figure** 3. Demonstration of 4D dose calculation: (a) each colored dot on the breathing curve represents a small fraction of time in delivery and corresponding to: (1) an individual 3D model/*iCT* image dataset representing the anatomy at that time, and (2) MU segments including the derived gantry angle, collimator angles, positions of each individual jaws and MLC leaves during that time. (b) MU Segments around time $t_x(i)$ and time $t_x(i+1)$ belong to model $t_x$. Similarly, MU segments around time $t_y(i)$ and time $t_y(i+1)$ belong to model $t_y$. All the MU segments on the same model/*iCT* will be assembled to a *SubPlan*. Then dose would be calculated for each *SubPlan* and deformed back onto Reference CT ($CT_{Ref}$). Finally, the deformed dose was then accumulated on the reference CT to generate the 4D dose.

## Results
### 1. Dosimetric comparison of interplay effect between IMRT and VMAT
**Figure 4** (a) plotted the %E/P results for the entire VMAT plan group and entire IMRT plan group. Overall, the %E/P for GTV mean dose was statistically different between IMRT and VMAT (p = 0.04). In average, it was lower for IMRT (0.5% ± 7.7%) than for VMAT (3.5% ± 5.0%) as shown in **Figure 4**(a). However, the differences in %E/P of GTV minimal dose between IMRT (2.2%±7.6%) and VMAT (5.1%±5.7%) was not statistically significant (p = 0.08).

**Figure 4**(b) plotted the standard deviation of the 4D GTV mean and minimal dose for both VMAT and IMRT groups. Even though the difference between the VMAT and IMRT is statistically significant, the magnitude of the standard deviation is small and within 2%.

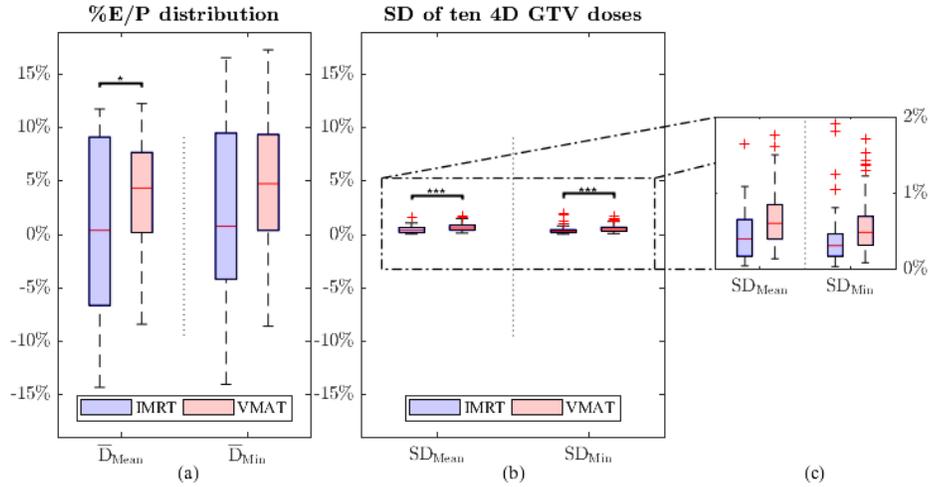

**Figure 4.** (a) Statistical boxplot of %E/P values for GTV mean and minimal dose comparing IMRT and VMAT plans, (b) standard deviation (SD) of 4D GTV dose with the same y-axis scale, and (c) the enlarged view for the dash line marked area. The black bars with stars were comparing data distribution between the IMRT group with the VMAT group. The p-value for the comparison was graphically represented by the number of stars. * represents $p<0.05$, ** represents $p<0.01$ and *** represents $p<0.001$.

**Figure 5** plotted the expected GTV minimal dose from the total 128 plans (64 IMRT and 64 VMAT) with different combinations of planning parameters for 4 tumor models. The expected GTV minimal dose was lower than the single fraction prescription dose in six IMRT plans (9.4 %) as marked by the blue arrows. Of the six IMRT plans, three were planned with 10 MV FFF, four planned with beam isocenter at the center-of-mass (COM) of the PTV (C-ISO), four planned with 3 mm margin and four had heterogeneous dose distribution in the PTV. In contrast, the expected GTV minimal dose in two VMAT plans (3.1 %), marked with red arrows, was lower than the single fraction prescription dose. Both of them were planned with 6 MV FFF beam with isocenter at Off-COM of the PTV and had homogeneous dose distribution in the PTV.

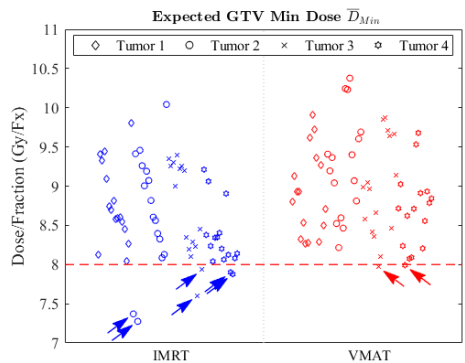

**Figure 5**. Expected GTV minimal dose $\bar{D}_{Min}$ for both IMRT (blue) and VMAT (red) techniques (red dash line indicates the single fraction prescription dose 8 Gy). Different tumor models were indicated with different marker shapes. There were 16 IMRT and 16 VMAT plans for each tumor model. The arrows marked the plans with $\bar{D}_{Min}$ being less than 8 Gy.

**Figure 6** plotted the percentage of plans under "no underdose" and "borderline no underdose" criteria in the IMRT and the VMAT group, respectively. The VMAT group showed a higher percentage than the IMRT group for both criteria.

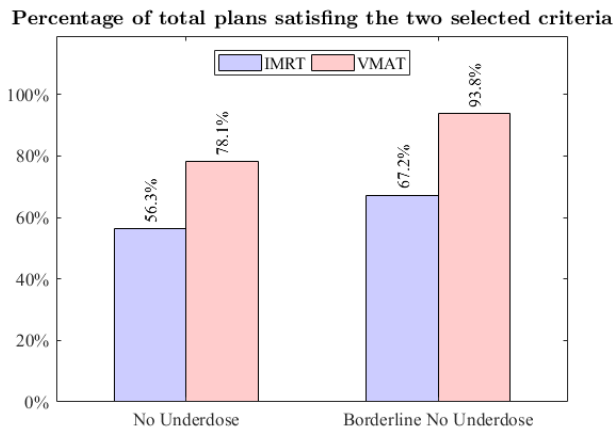

**Figure 6**. Percentage of the total number of IMRT or VMAT plans that satisfied the two criteria "no underdose" and "borderline no underdose".

## 2. Dosimetric evaluation of impact of planning parameters on interplay effect for IMRT and VMAT
### 2.A Expected dose versus planned dose (%E/P)

**Figure** 7 plotted the %E/P results for the subgroups of IMRT and VMAT in which one of the planning parameters was fixed (indicated by the labels on the X-axis). The *p* values of the comparisons were represented by stars.

For IMRT and VMAT plans with either homogeneous or heterogeneous dose distribution in the PTV, there is a significant difference between the two techniques in the %E/P values for both the GTV mean and minimal dose. Comparing IMRT to VMAT, in average, the %E/P values were higher for plans with homogeneous PTV dose distribution (5.5% ± 6.6% vs 3.2% ± 4.2% for mean dose with *p* = 0.011, and 6.5% ± 6.7% vs 3.3% ± 4.3% for minimal dose with *p* = 0.004), but lower for plans with heterogeneous PTV dose distribution -4.6% ± 5.1% vs 3.9% ± 5.8% with *p* = 0.000 for mean dose, and (-2.2% ± 5.7% vs 6.9% ± 6.3% with *p* = 0.000 for minimal dose).

For plans utilized a 5 mm ITV to PTV margin, the differences in the %E/P for GTV mean and minimal dose were also statistically significant between IMRT and VMAT. The average %E/P values were lower in IMRT than VMAT (-0.9% ± 8.2% v.s. 5.5% ± 4.9% with *p* = 0.008 for mean dose, and 0.4% ± 8.3% v.s. 7.4% ± 5.9% with *p* = 0.006 for minimal dose). For plans with the beam isocenter at the COM of PTV, only the %E/P values for GTV mean dose were statistically different comparing IMRT to VMAT (-0.4% ± 7.8% vs 3.7% ± 5.1% with *p* = 0.039). The differences in %E/P values of GTV mean and minimal dose between VMAT and IMRT in other subgroups, including 6 MV FFF, 10 MV FFF, 3 mm margin and O-ISO were not statistically significant.

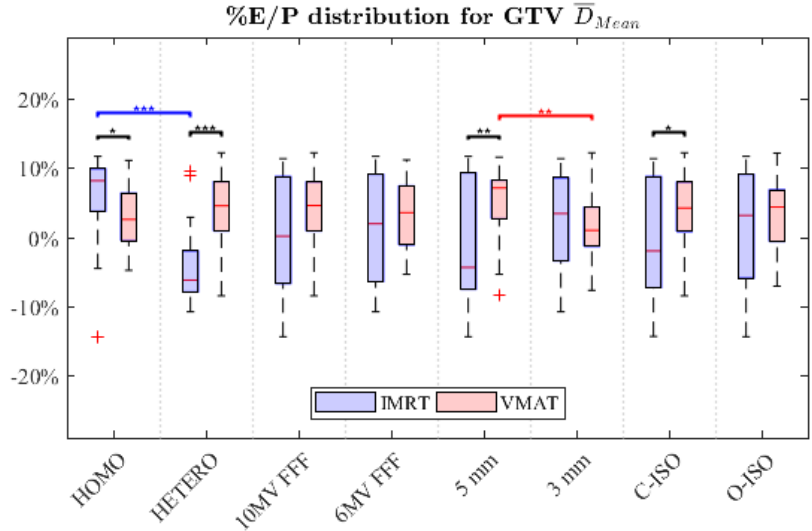

(a)

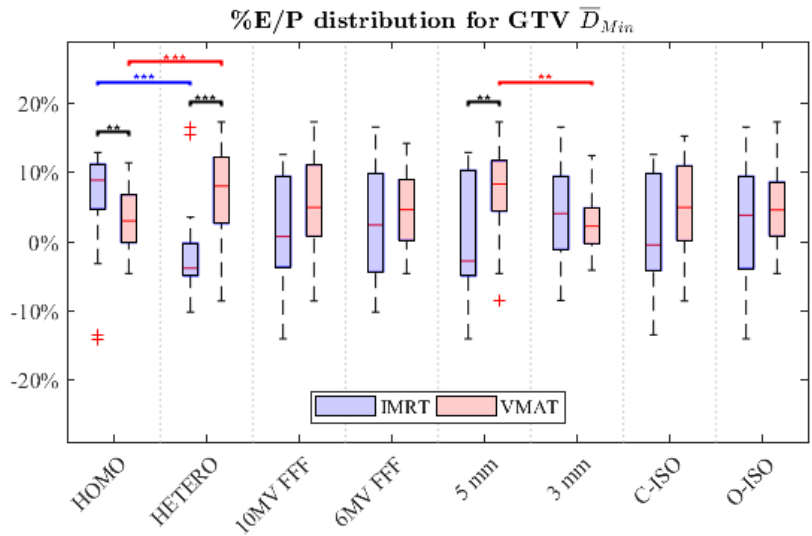

(b)

**Figure** 7. Statistical boxplot analysis of %E/P results between each IMRT and VMAT subgroups on (a) GTV mean dose and (b) GTV minimal dose. The black bars with stars represented comparisons between IMRT subgroups and VMAT subgroups. The blue bars with stars represented comparisons between two IMRT subgroups and the red bars represented comparisons between two VMAT subgroups. The p-value for the comparison was graphically represented by number of stars. * represents $p<0.05$, ** represents $p<0.01$ and *** represents $p<0.001$.

**Figure** 7 also demonstrated the dosimetric impact of the choice of beam energy, isocenter location, PTV margin, and dose distribution in PTV in either IMRT or VMAT on the motion interplay effect. For IMRT (comparisons indicated by the blue stars and bars), the %E/P values of GTV mean and minimal dose were significantly lower in the plans with heterogeneous dose distribution in PTV than those with homogeneous PTV dose (-4.6% ± 5.1% comparing to 5.5% ± 6.6% for mean dose with p <0.001, and -2.2% ± 5.7% comparing to 6.5%±6.7% for minimal dose with p <0.001). There was no statistically significant difference in %E/P values of GTV mean nor minimal dose between plans with different choices of beam energy, isocenter location, or PTV margin. For VMAT, the %E/P values for GTV minimal dose was significantly higher in plans with heterogeneous PTV dose than those with homogeneous PTV dose (6.9% ± 6.3% compared to 3.3% ± 4.3% with p <0.001), but no statistical difference for GTV mean dose. Also, using 5 mm PTV margin in planning resulted in higher %E/P values for both GTV mean and minimal dose than using 3 mm PTV margin (5.5% ± 4.9% comparing to 1.6% ± 4.4% for mean dose and 7.4% ± 5.9% comparing to 2.8% ± 4.4% for minimal dose). The choice of beam energy and isocenter location did not result in differences in %E/P values within each delivery technique for the GTV mean and minimal dose.

## 2.B Number of plans in each subgroup that satisfied the two criteria

**Figure 8**(a) demonstrated the number of plans that satisfied %E/P of the GTV minimal dose greater than zero (no underdose). Only in homogeneous PTV dose subgroup and 3 mm PTV margin subgroup, the ratio is higher for IMRT than for VMAT (87.5% vs 71.9% and 71.9% vs 68.8%, respectively).

**Figure 8**(b) demonstrated the number of plans that satisfied both %E/P of the GTV minimal dose greater than -3% and the GTV minimal dose greater than the prescription dose (borderline no underdose). For all subgroups, the number of IMRT plans is lower than VMAT plans.

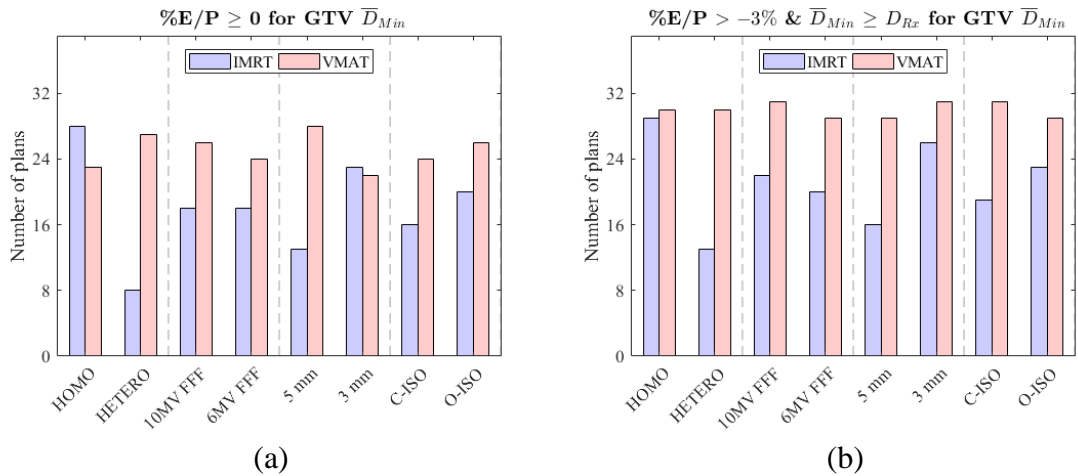

**Figure 8**. Total number of plans in IMRT or VMAT subgroups that satisfied the two selected criteria "no underdose" and "borderline no underdose"

**Figure 8** also showed the difference in the number of plans that satisfied the two predefined criteria when comparing the two paired subgroups within the same category for a given delivery technique (IMRT or VMAT). For IMRT, the percentage is higher for plans satisfied both criteria in the following subgroups (1) homogeneous PTV dose, (2) 3 mm PTV margin, and (3) beam isocenter at an off-COM point as compared to their corresponding counter-subgroups respectively. For VMAT, the percentage is higher for plans satisfied the "no underdose" condition for the following subgroups: (1) heterogeneous PTV dose, (2) beam energy of 10MV FFF, (3) 5 mm PTV margin, and (4) beam isocenter at an off-COM point, and higher for plans satisfied the "borderline no underdose" condition for the following subgroups: (1) 10MV FFF, (2) 3 mm PTV margin, and (3) beam isocenter at the COM of PTV, as compared to their corresponding counter-subgroups respectively.

## 3. Dosimetric evaluation of interplay effect for different tumor models

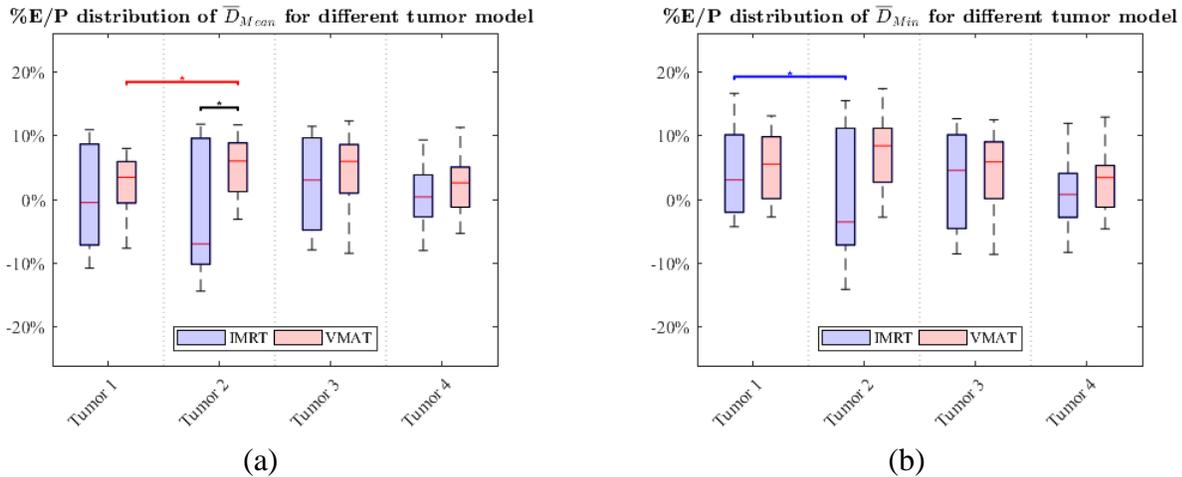

(a)                                              (b)

**Figure 9**. Statistical boxplot for %E/P values of (a) GTV mean dose and (b) GTV minimal dose for different tumor models. The black bar with star represented comparisons between IMRT and VMAT. The blue bars with stars represented comparison between two IMRT subgroups. The p-value for the comparison was graphically represented by number of stars. * represents *p<0.05*.

**Figure 9** plotted the %E/P values of GTV mean and minimal dose for all IMRT and VMAT plans for each of the four tumor models. Statistical comparisons were performed between paired groups with one tumor model parameter being different (namely between tumor 1 & tumor 2, between tumor 2 & tumor 3, and between tumor 3 & tumor 4). The difference for %E/P of GTV mean dose between IMRT and VMAT was statistically significant only for tumor model 2 (-1.5% ± 10.3% v.s. 5.1% ± 4.8%, with *p* = 0.049) as shown in **Figure 9**(a).

The comparison between plans on tumor model 1 and 2, where motion magnitudes were different (3 cm vs 1 cm), showed that the %E/P for GTV mean dose was significantly lower for tumor model 1 than model 2 (2.0% ± 4.7% v.s. 5.1% ± 4.8% with p =0.034) in VMAT planning, and %E/P for GTV minimal dose was significant higher for tumor model 1 than model 2 (3.9% ± 6.9% v.s. 0.6% ± 10.0% with *p* = 0.049) in IMRT planning. No other significant differences were observed in various comparisons among different tumor models.

Since tumor model 1 represented extreme scenario of 3 cm motion, further comparisons between plans on tumor model 1 and tumor model 2 were performed. As shown in **Figure 10**, comparison of the standard deviation of GTV minimal dose due to various initial position of tumor motion showed that both median value and range of variation of GTV minimal dose was significantly larger on plans for 3 cm motion than for 1 cm motion. This was true for IMRT or VMAT plans alone and all plans together. **Figure 11** showed comparisons of %E/P for GTV minimal dose between subgroups of both IMRT and VMAT plans on the 3cm and 1cm tumor motion. We observed that the %E/P for GTV minimal dose was significantly higher for tumors with 3cm motion in subgroups of IMRT plans where the dose distribution in PTV was heterogeneous or where a 5 mm PTV margin was used. While for subgroups of VMAT plans where the beam energy was 10 MV FFF, the %E/P for GTV minimal dose was significantly lower for tumor with 3cm motion.

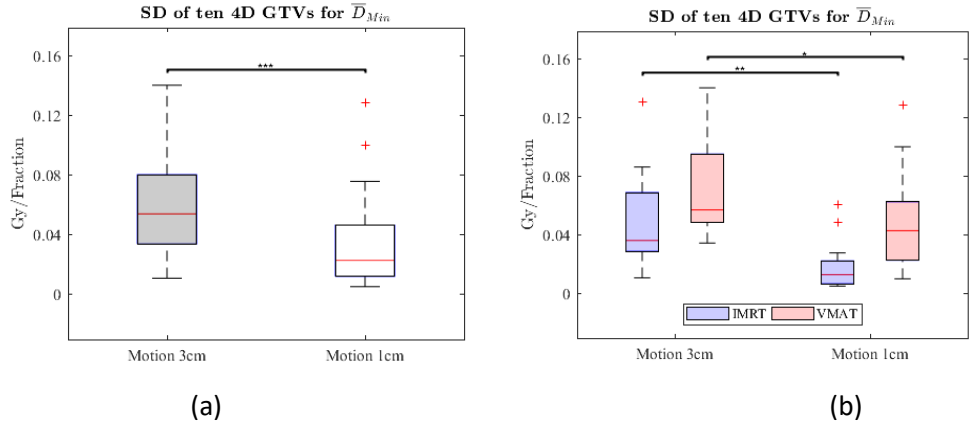

(a)                                                  (b)

**Figure 10.** Statistical boxplot analysis of standard deviation (SD) of ten minimal doses to GTV between the plans for 3 cm motion and 1 cm motion. (a) IMRT or VMAT plans. (b) IMRT and VMAT combined. The p-value for the comparison was graphically represented by number of stars, * represents p<0.05, ** represents p<0.01 and *** represents p<0.001.

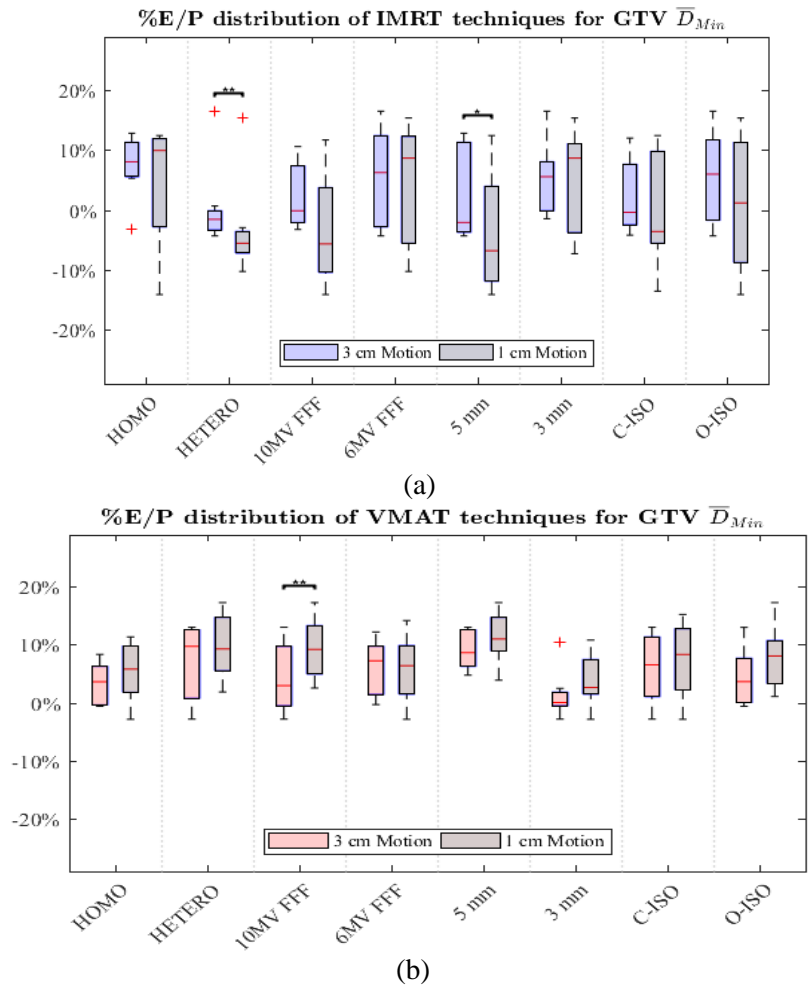

(a)

(b)

**Figure 11**. Statistical boxplot analysis of %E/P results for GTV minimal dose between 3 cm motion model and 1 cm motion model for subgroups of (a) IMRT plans and (b) VMAT plans. The p-value for the comparison was graphically represented by number of stars, * represents p<0.05, ** represents p<0.01.

# Discussion

The interplay between the simultaneous movement of the MLC and the tumor caused the delivered dose to deviate from the planned dose. As pointed out by Bortfeld [36], dose blurring due to breathing motion is independent of delivery technique and interplay effect is absent in conformal and static-compensator based IMRT technique. In the present study, we utilized a 4D dosimetry system based upon a virtual patient model to study the impact of different delivery techniques (step-and-shoot IMRT vs VMAT) and planning parameters on the motion interplay effect in SBRT. Due to the randomness of the initial position of tumor motion, the delivered dose/4D dose is random in nature. We computed the average and standard deviation of the dose-volume parameters from multiple delivered doses with a randomized initial position of tumor motion.

Our calculations have shown that even though the differences in uncertainties in the 4D GTV mean and minimal dose due to the random initial position of the tumor motion were statistically significant between IMRT and VMAT, the magnitudes are small (<2%) for all plans. Thus, in the SBRT setting, the expected dose-volume parameters may be used to quantify the dosimetric effect of motion interplay.

In our analysis, it should be noted that the dosimetry endpoints were dose-volume parameters such as mean and minimal dose. Since dose-volume parameters are often used in clinical practice for evaluating the plan quality, the direct point dose comparisons between static calculations and 4D calculations were not performed. The following discussion will be focused on the quantity %E/P and the expected GTV minimal dose.

## 1. GTV dose

A relatively larger variation of %E/P values was observed in our study, partially due to different combinations of delivery technique/planning parameter combinations investigated. Overall, the median value of %E/P for the GTV mean and minimal dose was above zero for both IMRT and VMAT as shown in **Figure 4**(a). However, it was shown in **Figure 6** that in about 44% of IMRT plans and 22% of VMAT plans, the %E/P value of GTV minimal dose was less than zero, which indicated that the delivered dose can be less than the planned dose due to motion interplay. It should be pointed out that SBRT planning is often characterized by a higher hot spot inside the PTV, therefore the planned dose to the ITV is generally higher than the prescription dose. We further compared the GTV minimal dose to the prescription dose for all the plans. As shown in **Figure 5**, the GTV minimal dose was lower than the single fraction prescription dose in six IMRT plans (9.4 % of the total plans) and two VMAT plans (3.1 % of the total plans). The expected GTV mean and minimal dose could also be higher due to motion interplay, but that was not a concern in this study. Thus, regardless of the delivery technique, the underdosing of GTV was observed. In the worst-case scenario, the GTV minimal dose can be 14.1% less than the ITV minimal dose and can be 9.0 % less than the prescription dose.

## 2. Choice of delivery technique / planning parameters to minimize motion interplay effect

**Figure 5** and **Figure 6** showed more IMRT plans demonstrated the underdosing of GTV due to motion interplay compared to VMAT plans. If a plan is considered acceptable taking into account motion interplay effect when the GTV dose satisfied the "borderline no underdose" condition (expected GTV minimal dose is no more than 3% lower than the planned ITV dose but greater than the prescription dose), a lower percentage for IMRT plans was also observed. These results suggest that IMRT planning might be more prone to motion interplay effects in SBRT in general.

One of the differences between VMAT and step-and-shoot IMRT delivery was the number of control points. In this study, we have purposely constrained the total number of segments used in the IMRT plan to be less than or equal to 20, which is much less than the number of control points in VMAT (~100). This study showed that restricting the number of control points in IMRT did not result in reduced motion interplay effect. One possible explanation of step-and-shoot IMRT is more sensitive to motion interplay could be the sharp transition of dose delivered between different segments. Thus, deviation from the planned position due to motion may introduce larger difference in dose.

This study also demonstrated the impact of the choice of other planning parameters used for either IMRT or VMAT. The comparisons of the distribution of %E/P for GTV minimal dose indicated the choice of desired dose distribution in PTV and PTV margin can have a significant impact. Based upon the number of plans in which GTV minimal dose satisfied the two criteria ("no underdose" and "borderline no underdose"), we concluded:

- In general, VMAT is to be recommended for planning. However, if a homogeneous dose distribution in the PTV is needed to reduce a hot spot on a critical organ that overlaps with the PTV, using IMRT is less likely to underdose the GTV due to motion interplay compared to VMAT.
- If the treatment machine is only capable of delivering static gantry angle IMRT, a 3 mm PTV margin can be used if there is accurate image guidance for patient setup, otherwise one should design a plan with a homogeneous dose distribution in PTV to reduce motion interplay effect.
- With VMAT, a 5 mm PTV margin or heterogeneous dose distribution in the PTV is preferred since the %E/P for GTV minimal dose and the number of plans with %E/P >0 was higher. However, if plans satisfying the "borderline no underdose" condition are acceptable, a 3 mm PTV margin may also be used.

The comparison of the distribution of %E/P values showed that the choice of beam energy and isocenter location has no significant impact on the motion interplay effect. However, **Figure 8** indicates more plans satisfied both criteria in Off-COM subgroup for IMRT. As for VMAT, more plans satisfied (1) both criteria in 10 MV FFF than in 6 MV FFF subgroup, (2) "no underdose" criteria in the off-COM than in COM subgroup, and (3) "borderline no underdose" criteria in COM than in off-COM subgroup. More studies are needed to investigate the impact of beam isocenter location and beam energy on the motion interplay effect. Note the off-COM point was only 5 mm away along each direction from the COM of the PTV in this study. A future study will be conducted on the motion interplay effect for the plans with beam isocenter at a point further away from the COM of the PTV.

### 3. Tumor models
The virtual patient model also enables us to study the motion interplay effect for different tumor sizes, locations, and motion amplitudes. The results indicated:

- Tumor model 2 which is an island tumor with 5 mm GTV and 1 cm motion, showed differences in the %E/P for GTV mean dose to be significantly higher with VMAT than with IMRT. Thus, using VMAT instead of IMRT will likely maintain the high mean dose on the GTV for an island lesion.
- The %E/P for GTV minimal dose was significantly higher for tumor model 1 than for tumor model 2 in IMRT planning, even though the motion magnitude was larger in tumor model 1 (3 cm motion vs 1 cm motion).

- The GTV size and motion amplitudes did not have a significant impact for the GTV minimal dose in VMAT plans. However, comparisons between tumor model 1 and tumor model 2 showed that the %E/P for GTV mean dose is significantly lower in VMAT planning for tumor model 1 that represented larger motion (3 cm).

Treating a lung tumor with 3 cm motion amplitude with SBRT should not be seen in routine clinical practice; however, it was simulated in this study to investigate the motion interplay effect in an extreme scenario. To further understand motion interplay effect for large motion, additional dosimetric analysis were performed to compare the plans for the 3 cm motion model and plans for the 1 cm motion model. In terms of the standard deviation of GTV minimal dose due to various initial position of tumor motion, both the medium value and the range of standard deviation were significantly larger for plans on 3 cm tumor motion (IMRT or VMAT, or combined), however, the magnitude remained very small (<2% of prescription dose). In terms of effect of planning parameters, we only found that the %E/P for GTV minimal dose was lower subgroup of VMAT plans with 10 MV FFF beams for tumor with 3 cm motion. In addition, among all the IMRT/VMAT plans for tumor model 1, the lowest %E/P for GTV minimal dose was -4.3% / -2.8% respectively and the GTV minimal dose in all the plans was greater than the prescription dose. Thus, motion interplay effect in SBRT for small tumors with large motion may not be significant.

**4. Limitations and future work**

This study focused on computing delivered dose to moving lung tumors in SBRT plans with different treatment delivery techniques (VMAT/ IMRT) and planning parameters based upon a virtual patient model. In order to isolate the motion interplay effect, the setup error and changes in patient geometry were not included in this study. Due to computational complexity, the simulation was performed on a limited number of tumor sizes, locations, and motion amplitudes. However, the four tumor models in this study were selected to represent the most clinically relevant scenarios as well as an extreme case. The tumor motion was assumed to be periodic, but the methodology developed in this study can be applied to an irregular breathing pattern, which will be studied future.

Another limitation was that the effect on the dose to the organs-at-risks (OARs) was not studied. This may only be important for central lesions where the OARs can be near the PTV. In this study, only one tumor model was centrally located, therefore we focused on the motion interplay effect on the target volume.

## Conclusion

This study utilized a dynamic virtual patient model to investigate the motion interplay effect in free-breathing lung SBRT plans for different tumor models. These plans were developed using step-and-shoot IMRT and VMAT techniques with different combinations of planning parameters. The dosimetric effect of motion interplay in lung SBRT was estimated by using the expected value of dose-volume parameters computed from different delivered doses with random initial positions of tumor motion. We found the expected GTV minimal dose can be less than the planned ITV minimal dose and even less than the prescription dose for both IMRT and VMAT plans. Secondly, the differences between the expected GTV dose and the planned ITV dose due to motion interplay depended on delivery technique and planning parameter which can be selected to minimize the interplay effect. In general, VMAT is to be recommended for planning. IMRT should be used if homogeneous dose distribution in the PTV was desired.

## Acknowledgement
The authors would like to thank Dr. Ping Xia for helpful discussion and support.

# Reference

[1] Timmerman R. Stereotactic Body Radiation Therapy for Inoperable Early Stage Lung Cancer. JAMA 2010;303:1070. https://doi.org/10.1001/jama.2010.261.

[2] Onishi H, Shirato H, Nagata Y, Hiraoka M, Fujino M, Gomi K, et al. Stereotactic body radiotherapy (SBRT) for operable Stage i non-small-cell lung cancer: Can SBRT be comparable to surgery? Int J Radiat Oncol Biol Phys 2011;81:1352–8. https://doi.org/10.1016/j.ijrobp.2009.07.1751.

[3] Videtic GMM, Donington J, Giuliani M, Heinzerling J, Karas TZ, Kelsey CR, et al. Stereotactic body radiation therapy for early-stage non-small cell lung cancer: Executive Summary of an ASTRO Evidence-Based Guideline. Pract Radiat Oncol 2017;7:295–301. https://doi.org/10.1016/j.prro.2017.04.014.

[4] Onishi H, Shirato H, Nagata Y, Hiraoka M, Fujino M, Gomi K, et al. Hypofractionated stereotactic radiotherapy (HypoFXSRT) for stage I non-small cell lung cancer: Updated results of 257 patients in a Japanese multi-institutional study. J Thorac Oncol 2007;2:S94–100. https://doi.org/10.1097/JTO.0b013e318074de34.

[5] Zhang B, Zhu F, Ma X, Tian Y, Cao D, Luo S, et al. Matched-pair comparisons of stereotactic body radiotherapy (SBRT) versus surgery for the treatment of early stage non-small cell lung cancer: A systematic review and meta-analysis. Radiother Oncol 2014;112:250–5. https://doi.org/10.1016/j.radonc.2014.08.031.

[6] Guerrero E, Ahmed M. The role of stereotactic ablative radiotherapy (SBRT) in the management of oligometastatic non small cell lung cancer. Lung Cancer 2016;92:22–8. https://doi.org/10.1016/j.lungcan.2015.11.015.

[7] Murray P, Franks K, Hanna GG. A systematic review of outcomes following stereotactic ablative radiotherapy in the treatment of early-stage primary lung cancer. Br J Radiol 2017;90. https://doi.org/10.1259/bjr.20160732.

[8] Videtic GMM, Stephans K, Reddy C, Gajdos S, Kolar M, Clouser E, et al. Intensity-modulated radiotherapy--based stereotactic body radiotherapy for medically inoperable early-stage lung cancer: excellent local control. Int J Radiat Oncol Biol Phys 2010;77:344–9.

[9] Matuszak MM, Yan D, Grills I, Martinez A. Clinical applications of volumetric modulated arc therapy. Int J Radiat Oncol Biol Phys 2010;77:608–16.

[10] Webb S. Motion effects in (intensity modulated) radiation therapy: A review. Phys Med Biol 2006;51:403–25. https://doi.org/10.1088/0031-9155/51/13/R23.

[11] Boda-Heggemann J, Knopf AC, Simeonova-Chergou A, Wertz H, Stieler F, Jahnke A, et al. Deep Inspiration Breath Hold - Based Radiation Therapy: A Clinical Review. Int J Radiat Oncol Biol Phys 2016;94:478–92. https://doi.org/10.1016/j.ijrobp.2015.11.049.

[12] Brandner ED, Chetty IJ, Giaddui TG, Xiao Y, Huq MS. Motion management strategies and technical issues associated with stereotactic body radiotherapy of thoracic and upper abdominal tumors: A review from NRG oncology. Med Phys 2017;44:2595–612. https://doi.org/10.1002/mp.12227.

[13] Molitoris JK, Diwanji T, Snider JW, Mossahebi S, Samanta S, Badiyan SN, et al. Advances in the use of motion management and image guidance in radiation therapy treatment for lung cancer. J Thorac Dis 2018;10:S2437–50. https://doi.org/10.21037/jtd.2018.01.155.

[14] Caillet V, Booth JT, Keall P. IGRT and motion management during lung SBRT delivery. Phys Medica 2017;44:113–22. https://doi.org/10.1016/j.ejmp.2017.06.006.

[15] Yu CX, Jaffray DA, Wong JW. The effects of intra-fraction organ motion on the delivery of dynamic intensity modulation. Phys Med Biol 1998;43:91–104. https://doi.org/10.1088/0031-

9155/43/1/006.
[16] Jiang SB, Pope C, Al Jarrah KM, Kung JH, Bortfeld T, Chen GTY. An experimental investigation on intra-fractional organ motion effects in lung IMRT treatments. Phys Med Biol 2003;48:1773.
[17] Duan J, Shen S, Fiveash JB, Popple RA, Brezovich IA. Dosimetric and radiobiological impact of dose fractionation on respiratory motion induced IMRT delivery errors: A volumetric dose measurement study. Med Phys 2006;33:1380–7. https://doi.org/10.1118/1.2192908.
[18] Schaefer M, Münter MW, Thilmann C, Sterzing F, Haering P, Combs SE, et al. Influence of intra-fractional breathing movement in step-and-shoot IMRT. Phys Med Biol 2004;49:N175.
[19] Court LE, Seco J, Lu X-QQ, Ebe K, Mayo C, Ionascu D, et al. Use of a realistic breathing lung phantom to evaluate dose delivery errors a. Med Phys 2010;37:5850–7. https://doi.org/10.1118/1.3496356.
[20] Berbeco RI, Pope CJ, Jiang SB. Measurement of the interplay effect in lung IMRT treatment using EDR2 films. J Appl Clin Med Phys 2006;7:33–42. https://doi.org/10.1120/jacmp.v7i4.2222.
[21] Litzenberg DW, Hadley SW, Tyagi N, Balter JM, Ten Haken RK, Chetty IJ, et al. Synchronized Dynamic Dose Reconstruction. Med Phys 2006;33:2138. https://doi.org/10.1118/1.2241321.
[22] Zhou S, Zhu X, Zhang M, Zheng D, Lei Y, Li S, et al. Estimation of internal organ motion-induced variance in radiation dose in non-gated radiotherapy. Phys Med Biol 2016;61:8157–79. https://doi.org/10.1088/0031-9155/61/23/8157.
[23] Ong C, Verbakel WFARAR, Cuijpers JP, Slotman BJ, Senan S. Dosimetric impact of interplay effect on rapidarc lung stereotactic treatment delivery. Int J Radiat Oncol Biol Phys 2011;79:305–11. https://doi.org/10.1016/j.ijrobp.2010.02.059.
[24] Rao M, Wu J, Cao D, Wong T, Mehta V, Shepard D, et al. Dosimetric impact of breathing motion in lung stereotactic body radiotherapy treatment using image-modulated radiotherapy and volumetric modulated arc therapy. Int J Radiat Oncol Biol Phys 2012;83:e251–6. https://doi.org/10.1016/j.ijrobp.2011.12.001.
[25] Edvardsson A, Nordström F, Ceberg C, Ceberg S. Motion induced interplay effects for VMAT radiotherapy. Phys Med Biol 2018;63. https://doi.org/10.1088/1361-6560/aab957.
[26] Tyler MK. Quantification of interplay and gradient effects for lung stereotactic ablative radiotherapy (SABR) treatments. J Appl Clin Med Phys 2016;17:158–66. https://doi.org/10.1120/jacmp.v17i1.5781.
[27] Zou W, Yin L, Shen J, Corradetti MN, Kirk M, Munbodh R, et al. Dynamic simulation of motion effects in IMAT lung SBRT. Radiat Oncol 2014;9:1–9. https://doi.org/10.1186/s13014-014-0225-3.
[28] Stambaugh C, Nelms BE, Dilling T, Stevens C, Latifi K, Zhang G, et al. Experimentally studied dynamic dose interplay does not meaningfully affect target dose in VMAT SBRT lung treatments. Med Phys 2013;40. https://doi.org/10.1118/1.4818255.
[29] Belec J, Clark BG. Monte Carlo calculation of VMAT and helical tomotherapy dose distributions for lung stereotactic treatments with intra-fraction motion. Phys Med Biol 2013;58:2807–21. https://doi.org/10.1088/0031-9155/58/9/2807.
[30] Zhao B, Yang Y, Li T, Li X, Heron DE, Huq MS. Dosimetric effect of intrafraction tumor motion in phase gated lung stereotactic body radiotherapy. Med Phys 2012;39:6629–37.
[31] Kubo K, Monzen H, Ishii K, Tamura M, Nakahara R, Kishimoto S, et al. Minimizing dose variation from the interplay effect in stereotactic radiation therapy using volumetric modulated arc therapy for lung cancer. J Appl Clin Med Phys 2018;19:121–7.

https://doi.org/10.1002/acm2.12264.

[32] Li X, Yang Y, Li T, Fallon K, Heron DE, Huq MS. Dosimetric effect of respiratory motion on volumetric-modulated arc therapy-based lung SBRT treatment delivered by TrueBeam machine with flattening filter-free beam. J Appl Clin Med Phys 2013;14:4370. https://doi.org/10.1120/jacmp.v14i6.4370.

[33] Giglioli FR, Clemente S, Esposito M, Fiandra C, Marino C, Russo S, et al. Frontiers in planning optimization for lung SBRT. Phys Medica 2017;44:163–70.

[34] Guo B, George Xu X, Shi C. Real time 4D IMRT treatment planning based on a dynamic virtual patient model: Proof of concept. Med Phys 2011;38:2639–50. https://doi.org/10.1118/1.3578927.

[35] Bezjak A, Bradley J, Gaspar L, Timmerman RD. Seamless Phase I / II Study of Stereotactic Lung Radiotherapy (SBRT) for Early Stage, Centrally Located, Non-Small Cell Lung Cancer (NSCLC) in Medically Inoperable Patients. 2015.

[36] Bortfeld T, Jokivarsi K, Goitein M, Kung J, Jiang SB. Effects of intra-fraction motion on IMRT dose delivery: Statistical analysis and simulation. Phys Med Biol 2002;47:2203–20. https://doi.org/10.1088/0031-9155/47/13/302.

[37] Brock KK, Mutic S, McNutt TR, Li H, Kessler ML. Use of image registration and fusion algorithms and techniques in radiotherapy: Report of the AAPM Radiation Therapy Committee Task Group No. 132. Med Phys 2017;44:e43--e76.

[38] Sarrut D, Baudier T, Ayadi M, Tanguy R, Rit S. Deformable image registration applied to lung SBRT: Usefulness and limitations. Phys Medica 2017;44:108–12. https://doi.org/10.1016/j.ejmp.2017.09.121.